\documentclass[sigconf]{acmart}

\AtBeginDocument{%
  \providecommand\BibTeX{{%
    \normalfont B\kern-0.5em{\scshape i\kern-0.25em b}\kern-0.8em\TeX}}}

 \setcopyright{acmcopyright}

\copyrightyear{2022} 
\acmYear{2022} 
\setcopyright{acmlicensed}\acmConference[Chinese CHI 2022]{The Tenth International Symposium of Chinese CHI}{October 22--23, 2022}{Guangzhou, China and Online, China}
\acmBooktitle{The Tenth International Symposium of Chinese CHI (Chinese CHI 2022), October 22--23, 2022, Guangzhou, China and Online, China}
\acmPrice{15.00}
\acmDOI{10.1145/3565698.3565774}
\acmISBN{978-1-4503-9869-5/22/10}

\usepackage{xcolor}
\newcommand{\es}[1]{\textcolor{black}{#1}} 
\begin{document}


\title[Older Adults' Perceptions and Challenges with AI-enabled Everyday Technologies]{Understanding Older Adults' Perceptions and Challenges in Using AI-enabled Everyday Technologies}


\author{Esha Shandilya}
\affiliation{%
  \institution{School of Information} \institution{Rochester Institute of Technology}
  \city{Rochester}
  \state{NY}
  \country{USA}
}
\email{es4524@rit.edu}

\author{Mingming Fan}
\authornote{Corresponding Author}
\orcid{0000-0002-0356-4712}
\affiliation{
  \institution{Computational Media and Arts Thrust}
 \institution{The Hong Kong University of Science and Technology (Guangzhou)}
  \city{Guangzhou}
  \country{China}
}
\affiliation{
\institution{Division of Integrative Systems and Design}
  \institution{The Hong Kong University of Science and Technology}
  \city{Hong Kong SAR}
  \country{China}
}
\email{mingmingfan@ust.hk}


\begin{abstract}
Artificial intelligence (AI)-enabled everyday technologies could help address age-related challenges like physical impairments and cognitive decline. While recent research studied older adults' experiences with specific AI-enabled products (e.g., conversational agents and assistive robots), it remains unknown how older adults perceive and experience current AI-enabled everyday technologies in general, which could impact their adoption of future AI-enabled products. We conducted a survey study (N=41) and semi-structured interviews (N=15) with older adults to understand their experiences and perceptions of AI. We found that older adults were enthusiastic about learning and using AI-enabled products, but they lacked learning avenues. Additionally, they worried when AI-enabled products outwitted their expectations, intruded on their privacy, or impacted their decision-making skills. Therefore, they held mixed views towards AI-enabled products such as AI, an aid, or an adversary. We conclude with design recommendations that make older adults feel inclusive, secure, and in control of their interactions with AI-enabled products.
\end{abstract}




\maketitle

\section{Introduction}
\label{sec:intro}
Artificial Intelligence (AI)-enabled products, such as online shopping and autonomous driving, are becoming increasingly integrated into daily lives that people are often unaware of their presence and potential impact on daily lives~\cite{ARM:survey,zhang2019artificial}. 
Recent research began to understand people's experiences and attitudes toward AI and showed that age may affect their attitudes and experiences with AI~\cite{ARM:survey,Bristows:survey,zhang2019artificial}. While younger adults tend to encounter AI products more often and are generally positive about AI, older adults aged 60 years or above~\cite{CDC,UN,WHO} tend to have relatively less experience with AI products~\cite{Bristows:survey} and may face challenges adapting and using AI products~\cite{atkinson2016breaking}. 
As AI could potentially help older adults deal with age-related issues such as physical impairments \cite{jung2018potential}, cognitive decline~\cite{chan2020biosignal}, and emotional isolation~\cite{zuckerman2020companionship}, it is crucial to understand older adults' personal encounters and experiences with AI to better inform the design of AI technologies for this population.

Recent research began to explore older adults' experiences with specific AI-enabled products, such as conversational agents (e.g., Alexa, Google Home)~\cite{beaney2020alexa,bonilla2020older,kowalski2019older,lopatovska2018personification,o2020voice,pradhan2019phantom,trajkova2020alexa}, smart appliances (e.g., smart vacuum cleaner)~\cite{kwan2019examining, ng2012older,smarr2011understanding,ramalingam2020human,pearce2012robotics}, and assistive robots~\cite{kowalski2019older,pradhan2020understanding,knowles2019hci,wu2018understanding}.   
While experiences with specific AI-enabled products are imperative to improve these products, the general perceptions and understanding of AI-enabled products affect older adults' acceptance and expectations of current and future AI-enabled technologies in general. Furthermore, it remains largely unknown i) how older adults perceive the term ``AI" and expect AI-enabled technologies to work, ii) their knowledge of AI and trust in AI, and iii) the reasons why they might have specific (mis)perceptions about AI. 

In this research, we aim to understand qualitatively older adults' knowledge, experiences and (mis)perceptions of using AI-enabled products by exploring the following research questions (RQs):
\begin{itemize}
    \item \textbf{RQ1:} What are older adults’ personal experiences with AI-enabled products?
    \item \textbf{RQ2:} Do older adults face any challenges in using AI-enabled products? If yes, what are they?
    \item \textbf{RQ3:} What are older adults' perceptions of the AI technology in AI-enabled products?
\end{itemize}

We first conducted an online survey study with $41$ participants to screen and recruit older adults (60 or above) who have experience in using AI-enabled products. Through our survey study, we found that despite older adults' limited knowledge of AI, they used a variety of AI-enabled products, though some products like customer-service chatbots, recommendation features, and voice-assistants were used less frequently than other ones like search engines and navigational maps that use AI in the background. To gain a deeper understanding of older adults' responses to the survey questions, we recruited $15$ survey respondents who had some experience of using AI-enabled products for interviews. Our interview findings show that older adults' limited awareness of AI might be attributed to the lack of a tech-savvy community and the negative portrayal of AI in news and media that discouraged them from exploring AI-enabled products.
Moreover, other adversarial behaviors or consequences of using AI-enabled products such as AI stealing their data, creating smarter scams, and owning their data further made them hesitant to explore AI-enabled products. Lastly, we discuss design considerations based on older adults' experiences and challenges in using AI-enabled products.
In sum, we make the following contributions in this work: 
\begin{itemize}
    \item An initial understanding of older adults' personal experiences and challenges in using AI-enabled products;
    \item A qualitative understanding of older adults' general perceptions and understanding of the AI used in AI-enabled products;
    \item Design considerations to make AI-enabled products more acceptable and accessible to older adults.
\end{itemize}

\section{Background and Related Work}
\label{sec:background}
\subsection{Older Adults and Technology Use}
\label{subsec:OAandTech}
Technological products facilitate us from ordering food, shopping, managing health care to financial and social security services \cite{czaja1993computer}, and weaves into the lives of all, including older adults, aged 60 or above~\cite{CDC,UN,WHO}. Keeping pace with the changing technology becomes crucial for oneself \cite{leahy2009digital}. Adapting to the rapidly changing technology is often left to an individual’s technological capabilities and skill-sets~\cite{atkinson2016breaking}. Melenhorst et al.~\cite{melenhorst2006older} argue that specifically older adults deter to adopt new advancements in technology not because of usability issues or cost, but because of their lack of awareness about the benefits or value the technology could provide. Older adults’ motivation to use and invest their energy in new technology depends on their knowledge of its perceived advantages and worth~\cite{czaja1993computer,melenhorst2006older}. 

Digital literacy is of paramount importance for an individual to determine the perceived use and value of a technology~\cite{atkinson2016breaking}. According to Leahy and Dolan~\cite{leahy2009digital}, digital literacy means individuals’ ability to discern the usefulness of a particular technology and their comfort in using it for various purposes; it applies to anyone from a non-technological background, including older adults. However, to attain this skill and proficiency, technology should be usable and accessible to everyone~\cite{coleman2010engaging}. Older adults experience anxiety, fear, lack of control, and unawareness of existent services while using any technological product \cite{atkinson2016breaking}. Additionally, they also encounter ergonomic\cite{kurniawan2006study} and usability issues \cite{braun2013obstacles,luger2014older,de2016study} in technological products like online banking~\cite{gatsou2017seniors},mobile phones~\cite{kurniawan2006study,kurniawan2008older}, tablet computers~\cite{vaportzis2017older}, and websites~\cite{braun2013obstacles,luger2014older,de2016study}. 

To summarize, past work suggest that an older adult's motivation to use and adapt an advancing technology depends on three factors: i) technology's perceived usefulness and potential~\cite{czaja1993computer,melenhorst2006older}, ii) digital literacy to experience technology's benefits~\cite{atkinson2016breaking}, and iii) personal apprehensions in using digital technologies~\cite{atkinson2016breaking}. However, these factors were mostly derived based on non-AI technologies. As recent advancements have been continuing introducing AI into various technological products, we are yet to understand how and in what forms older adults' motivations manifest into (un)adoption and usage of AI.


\subsection{Older Adults and AI}
\label{subsec:OAandAI}
AI technology has become part and parcel of our daily lives \cite{long2019designing}. AI has revolutionized many sectors like the e-commerce industry \cite{soni2020emerging,ricci2002recommendation}, from providing personalized recommendations while shopping \cite{ricci2002recommendation}, virtual shopping assistants to help us with queries \cite{soni2020emerging}, to detecting frauds \cite{nanduri2020microsoft}. Additionally, AI is extensively used in healthcare sectors for early disease diagnosis \cite{shen2019artificial}, in social media to enhance users’ engagement \cite{furman2019ai}. Many times, users are not aware that they interact with products that use AI, which affects their perception and may create false expectations from AI systems \cite{ARM:survey,long2020ai}. Therefore, Long and Magerko \cite{long2020ai} emphasize the importance of AI literacy for users and that digital literacy is the prerequisite for it.


AI has been used
in enhancing the lives of older adults~\cite{beaney2020alexa,bonilla2020older,felberbaum2020enhancing,knowles2019hci,pearce2012robotics}. Older adults experience various age-related issues such as loneliness, functional decline impacting their physical and mental abilities hindering them from performing their daily life activities comfortably~\cite{d2019mini}. Consequently, most of the research studies aim at resolving/assisting older adults with age-related issues through \textit{specific} AI-enabled technological interventions~\cite{bruun2016motivating,d2019mini,de2010use}. For example, how conversational agents like Alexa, smart appliances like a smart cleaner~\cite{ramalingam2020human}, or assistive robots augment their physical or cognitive abilities~\cite{chan2020biosignal,jung2018potential,hagiya2016typing,mostajeran2020augmented,ng2012older,wu2018understanding}, alleviate the social isolation felt by older adults~\cite{cyarto2016active,lopatovska2018personification}, and their feelings towards it~\cite{bonilla2020older,o2020voice,pradhan2019phantom,smarr2011understanding,trajkova2020alexa}. However, we still do not know older adults' \textit{general} perceptions, experiences, and concerns related to AI-enabled technologies, which could ultimately play a significant role in setting their expectations of AI products and their decisions to embrace them. 

A few recent studies~\cite{ARM:survey,Bristows:survey,shandwick2016ai,zhang2019artificial} studied people’s awareness, applications, challenges, benefits, and drawbacks of AI. 
Holder et al.~\cite{Bristows:survey} conductd a survey study with more than 2000 UK respondents and found that more than 74\% of the older adults aged 55 years and above have limited AI knowledge and limited exposure to AI products as compared to younger adults~\cite{Bristows:survey}. However, none of these earlier studies~\cite{ARM:survey, Bristows:survey, zhang2019artificial} were oriented towards studying older adults' perceptions of \textit{general} AI technology, experiences and needs in using an AI-enabled product(s) effectively. Further, these studies have not focused on how older adults' perceptions impact their usage or vice-versa. 
Therefore, we take a step further to understand older adult's perceptions of AI, experiences in using AI-enabled products, and their expectations from such products through qualitative research methods by focusing on \textit{general} AI technology.





\section{Online Survey Study}
\label{sec:survey}
To understand older adults’ experiences in using AI-enabled products and their perceptions about AI technology, we first conducted an online survey \es{to recruit} participants aged 60 or above, a common age threshold for older adults used by United Nation, World Health Organization and prior research~\cite{CDC,UN,WHO}. In this section, we describe our survey methodology and survey findings that informed the follow-up interview study. We further discuss the interview study in Section~\ref{sec:interviews} and interview findings in Section~\ref{sec:interviewfindings}.
All studies were approved by IRB at our institution.

\subsection{Materials and Procedure}
\label{subsubsec:materials}
\es{To identify older adults aged 60 or above with experience in using AI-enabled products, 
we designed 7 survey questions (6 close-ended and 1 optional open-ended). 
We did not provide a formal definition of AI in the survey for two reasons. First, we would like to keep it open so as to elicit older adults' understanding of AI. Second, a formal definition of AI might be too technical to be understood by all potential older adult participants. Instead, we used $11$ AI-enabled services as examples to probe their perceptions and understanding of AI: i) customer service chatbots, ii) email spam filter, iii) navigational maps like Google Maps, iv) search engines like Google, v) video recommendation features on YouTube, Netflix, vi) online shopping recommendation features on Amazon, vii) personal finance like mobile-check deposit, viii) ride-sharing applications like Uber, ix) robotic vacuum e.g. iRobot, x) social-network applications like Facebook, and xi) voice-assistants like Alexa.}
\es{Our questions were centered around collecting: i) demographic information \textit{(older adults' age and gender)}, ii) frequency of AI-enabled products used by older adults, iii) their perceived AI knowledge level, and iv) their sources, if any, used to learn AI. 
The survey also allowed participants to sign up for a follow-up interview by providing their email-ids.
}

We advertised our survey on social advertisement platforms (e.g., Craigslist, Facebook, Reddit) to target older adults from the North-American region. We also contacted local senior communities and adopted the snowball sampling technique during a six-week period to collect responses. The advertisements directly defined our target participant group---older adults--- and the purpose of the study. Participants completed the questionnaire in their own time and entered a raffle to win a $\$30$ award.

\subsubsection{Survey Participants}
\label{subsubsec:surveyparticipants}
We received $41$ survey responses in total but removed $6$ because they did not acknowledge being an older adult (60 years or above). All the survey responses gathered were from North-American countries.
Among the remaining $35$, $27$ reported their demographic information: gender (16 females, 10 males, and 1 preferred not to say) and age (6 between 60 and 69, 13 between 70 and 79, and 8 were 80 or above).
A total of $23$ respondents signed up to participate in a follow-up interview, and we eventually recruited $15$ of them who reported using at least  two AI-enabled products for the interviews.

Next, we briefly report our survey findings to provide an overview on: i) self-reported knowledge level of AI, ii) AI learning resources used, and iii) frequency of use of AI-enabled products by the survey participants.


\subsection{Survey Results}
\label{subsubsec:surveyresults}
We asked the respondents to report their perceived AI knowledge level by choosing a score on a scale of $1$ to $5$ (where 1 = `\textit{I am not aware of anything about AI.}' and 5 = `\textit{I am very knowledgeable about AI.}'), as shown in Table~\ref{tab:knowledgelevel}. Most of our survey participants, 20 out of 35 participants, reported having some knowledge about AI. 
We then presented a list of AI learning sources as shown in Table~\ref{tab:learningsources} with an ``other'' option to the respondents to select or provide sources that they used to learn anything about AI. The most used learning source is ``Internet search'' followed by ``Friends/Family members'' and ``Newspaper''; apart from the listed source, older adults also reported other learning sources like radio and news channels.

Our survey findings showed that participants used a variety of AI-enabled products. The most frequently used AI-enabled products are search engines like Google, 27 out of 35 participants use it, followed by navigational applications like Google Maps (25 out of 35) and email spam filter (16 out of 35) and online shopping recommendation feature (16 out of 35). However, respondents 
occasionally
used customer service chatbots (7 out of 35), recommendation systems on video platforms such as Netflix (11 out of 35) and online shopping sites like Amazon (10 out of 35), voice-assistants like Alexa (11 out of 35), and none of our participants (0 out of 35) used Robotic Vacuums.

\begin{table}[htb!]
\caption{This survey data shows the self-reported AI knowledge levels (i.e., scores and descriptions) among older adults.} 
\label{tab:knowledgelevel}
\Description{A table \label{tab:knowledgelevel} demonstrates Self-reported Score, Perceived AI Knowledge Level, and Number of Survey Participants }
\begin{tabular}{p{0.7cm}|p{5.0cm}|p{1.5cm}}
\toprule
\textbf{Score}  & \textbf{Perceived AI Knowledge Level}                                        & \textbf{No. of responses}\\
\bottomrule
1 & I am not aware of anything about AI.                      & 0   (0.00\%)          \\
2 & I am aware of AI but do not have much knowledge about it. & 12  (34.28\%)             \\
3 & I have some knowledge about AI.                           & 20  (57.14\%)          \\
4 & I am knowledgeable about AI.                              & 3   (8.5\%)        \\
5 & I am very knowledgeable about AI.                         & 0   (0.00\%)       \\
\bottomrule
\end{tabular}
%
\end{table}

\begin{table}[]
\caption{This table lists the AI learning sources that older adults use to learn about AI; the respondents specified radio and news channels as other learning sources. Participants were allowed to select multiple learning sources that they used to learn AI.} 
\label{tab:learningsources}
\Description{A table \label{tab:learningsources} demonstrates AI Learning Sources and Number of Survey Responses }
\begin{tabular}{p{6.2cm}|p{1.8cm}}
\toprule
\textbf{{AI Learning Sources}}                             & \textbf{{No. of Responses}} \\ \bottomrule
Internet search (e.g., Google)                                          & 23 (18.40\%)                              \\
\hline
Friends or family members                                               & 22 (17.60\%)                              \\
\hline
Newspapers                                                                  & 22 (17.60\%)                          \\
\hline
Online video sharing platforms (e.g., Youtube)                          & 11 (8.80\%)                               \\
\hline
TV channels                                                             & 11 (8.80\%)                               \\
\hline
Movies                                                                  & 10 (8.00\%)                                       \\
\hline
Social media (e.g., Facebook, Twitter)                                  & 10 (8.00\%)                                       \\
\hline
Others (please specify)                                                 & 8  (6.40\%)                                      \\
\hline
Academic articles (e.g., computer science journal or conference papers) & 6  (4.80\%)                                      \\
\hline
Massive open online course (e.g., Coursera, edX)                        & 2  (1.60\%)                                    \\
\bottomrule
\end{tabular}
%
\end{table}

\subsection{Summary of Survey Findings}
\label{subsubsec:surveysummary}
Our survey revealed that most of our survey respondents prefer using various AI products irrespective of their
age and have some knowledge of AI. Moreover, respondents leverage search engines,
newspapers, and depend on families
and friends to learn about AI. However, we do not yet know the reasons for older adults' limited knowledge of AI
technology and their sparse use of certain AI products.

The survey study helped us in identifying the older adults who use AI-enabled products but, we still did not have a
deep understanding of i) the experiences of older adults in using AI-enabled products, ii) if they face any difficulties in
using AI-enabled products (and what), iii) how their experiences with AI-enabled products inform their perception of
AI technology, and iv) their expectations from the AI technology. Therefore, to answer those questions and understand the
experiences of older adults in detail, we conducted online semi-structured interviews.

\section{Online Interview Study}
\label{sec:interviews}
\subsection{Materials and Procedure}
\label{subsubsec:interviewmaterials}
To gain qualitative insights into the experiences of older adults in using AI-enabled products, and their feelings and attitudes towards AI technology, we conducted online interviews. Our semi-structured interviews were designed to learn participants': i) understanding of AI technology, ii) feelings towards AI and reasons for those, iii) experiences using AI-enabled product(s), iv) barriers in embracing AI technology, and v) perspectives on making AI more accessible and usable for older adults, and follow-up questions on participant's survey responses.

We conducted semi-structured interviews on Zoom. The sessions were recorded for data analysis with the consent of the participants. The interviews were scheduled for approximately an hour. The mean recording time from the interviews was $72$ minutes ($std. = 10.52, min = 54, max = 91$).

\subsection{Interview Participants}
\label{subsubsec:interviewparticipants}
We interviewed $15$ participants (8 females and 7 males), and their demographic details (age range, gender, perceived AI knowledge level, and AI-enabled products used by them) are listed in the Table~\ref{tab:demographic}. 
We also list the most frequently used AI-enabled products reported by older adults in the survey study in Table~\ref{tab:demographic}.

\begin{table*}[t]
\caption{Interview participants' demographic information. We encode self-reported AI knowledge Level as 1:  I am not aware of anything about AI, 2: I am aware of AI but do not have much knowledge about it, 3: I have some knowledge about AI, 4 - I am knowledgeable about AI, 5 - I am very knowledgeable about AI. In this table, we also show the most frequently used AI-enabled products self-reported by older adults in the survey.}  
\label{tab:demographic}
\Description{A table labeled \label{tab:demographic} demonstrates Gender, Age Range, Perceived AI Knowledge Level, and Self-reported AI-enabled products Used} 

\begin{tabular}{p{0.5cm}|p{1.0cm}|p{1.5cm}|p{2.4cm}|p{10cm}}
\toprule
\textbf{ID} &\textbf{Gender} & \textbf{Age Range}  & \textbf{Perceived AI Knowledge Level} & \textbf{Self-reported AI-enabled Products Used}\\

\midrule
P1                                          & Male                              & 80-84      & 2    &    Customer Service Chatbots, Search Engine, Navigational Application, and Voice Assistant   \\
 \hline
P2                                          & Male                              & 75-79      & 3  & Customer Service Chatbots, Email Spam Filter, Search Engine, Navigational Application, Online Shopping Recommendation, Personal Finance, and Voice Assistant            \\
 \hline
P3                                        & Male                              & 65-69      & 2    & Email Spam Filter, Search Engine, Online Shopping Recommendation, Personal Finance, and Social Network             \\
 \hline
P4                                          & Female                            & 80-84      & 2   &  Email Spam Filter and Search Engine                   \\
 \hline
P5                                          & Female                            & 70-74      & 3  & Customer Service Chatbots, Email Spam Filter, Search Engine, and Navigational Application                     \\
 \hline
P6                                          & Male                              & 65-69      & 2  & Email Spam Filter, Search Engine, Navigational Application, Personal Finance, Video Recommendation, and Voice Assistant              \\
 \hline
P7                                          & Female                            & 75-79      & 3   & Customer Service Chatbots, Search Engine, Social Network, Online Shopping Recommendation, Personal Finance, and Voice Assistant       \\
 \hline
P8                                          & Female                            & >=85  & 2        & Customer Service Chatbots, Search Engine, Social Network, and Video Recommendation                      \\
 \hline
P9                                          & Male                              & >=85  & 2  & Search Engine, Navigational Application, Personal Finance, Social Network, and Video Recommendation, Voice Assistant               \\
 \hline
P10                                         & Male                              & 75-79     & 3             & Customer Service Chatbots, Email Spam Filter, Search Engine, Navigational Application, Personal Finance, and Social Network        \\
 \hline
P11                                         & Female                            & 65-69     & 3.          & Search Engine, Navigational Application, Personal Finance, and Social Network            \\
 \hline
P12                                         & Male                              & 60-64     & 2  & Customer Service Chatbots, Search Engine, Navigational Application, and Video Recommendation                 \\
 \hline
P13                                         & Female                            & 75-79     & 3         & Email Spam Filter, Search Engine, Navigational Application, Online Shopping Recommendation, and Video Recommendation   \\
 \hline
P14                                         & Female                            & 70-74     & 3       & Search Engine, Social Network, and Voice Assistant   \\
 \hline
P15                                         & Female                            & 75-79     & 3        & Email Spam Filter, Search Engine, and Online Shopping Recommendation     \\
\bottomrule
\end{tabular}
\end{table*}

\subsection{Analysis}
\label{sec:analysis}
We performed thematic analysis approach~\cite{braun2006using} to identify themes and relationships among qualitative data (interview data).
We transcribed the interviews and then familiarized ourselves with the transcripts. We then individually and iteratively coded the data to extract the key insights from the transcripts. We met and discussed the codes through weekly meetings to gain a consensus on our interpretations. We then identified the underlying sub-themes emerging from the codes and clustered the sub-themes into themes. The themes, sub-themes, and representative quotes were used to present our findings in the next section. 

\section{Online Interview Findings}
\label{sec:interviewfindings}
\es{Through our thematic analysis, we answer our three research questions \textit{(RQ1, RQ2, and RQ3)} under three themes: i) Older Adults' Experiences in Using AI-enabled Products, ii) Challenges in Using AI-enabled Products, and iii) Older Adults' Perception of AI Technology Based on their Personal Experiences with AI-enabled Products. We now discuss the narrative of each theme in detail using quotes from participants.}

\subsection{Older Adults' Experiences in Using AI-enabled Products (RQ1)}
This section explores how older adults adopted and learned about various AI-enabled products and their unique insights into how such products changed their interaction behavior. In Section~\ref{sec:survey}, we discussed that older adults use various AI-enabled products. However, we do not know how they gain knowledge about a product and learn how to use a product. Moreover, we also discuss how they interact with AI-enabled products in their day to day lives. From the survey study, Section~\ref{sec:survey}, we found that the most frequently used products were search engines like Google Search, navigational applications like Google Maps, and email spam filters. During the interview study, when we followed-up with interview participants on their survey responses, and asked them their experiences with the most used products (search engines, navigational maps, and email spam filter), they said that using these products has become their second nature. However, interview participants were more excited to share their experiences about other AI-enabled products like recommendation systems on video and shopping platforms, and voice-assistants. Therefore, we share their insights specific to those products.

\subsubsection{Adopting AI-enabled Products}
\subsubsection*{\textbullet{} Serendipitous Discovery}
Participants (P1, P5, P8, P9, and P13) adopted an AI-enabled product not because of their active choices but rather accidentally discovering those. For instance, they either received such products as gifts or bought the AI-enabled products because other conventional non-AI products were out of stock. 
\begin{quote}
    P9: ``... I have a grandson who works at Amazon and gave me Alexa.''
\end{quote}
\begin{quote}
    P5: ``... I went to the store and that's the only ones they had in stock because of the pandemic so many appliances were not being built because they couldn't put people in factories close together to build them. So I would not have bought a fancy smart washer and dryer normally, but that was the only one that I could find and I needed a washer.''
\end{quote}

\subsubsection*{\textbullet{} Informed Choices}
Several interviewees (P2, P3, P4, P6, P7, P10-P12, P14, and P15) also made informed decisions about their purchases. Major sources they gained knowledge about products were consumer reports or recommendations from friends and relatives.
    \begin{quote}
        P4: ``... Most of the information on my products I either get from consumer reports, or I will read uh evaluations on Amazon occasionally...So, for like the automatic vacuum the Roomba I think it is okay I have no interest in that at all okay, and I don't have animals so I don't have to worry.''
    \end{quote}
        \begin{quote}
        P14: ``... I'm more likely to enlarge what I've been doing try a new thing if I've had seen somebody do it or somebody said hey did you ever try yes okay um so that makes me curious to use a particular product or at least try it yeah. and if I'm willing to buy a particular product I talk to my friends to who's got what um particularly the ones whom I respect.''
    \end{quote}

Additionally, witnessing an AI-enabled product's usefulness through some personal encounters was another motivation for adoption.     

\begin{quote}
        P4: ``... I was in my son's house and we were talking about Alexa in his kitchen. When he used the word Alexa, it (the device) picked up and said what can I do for you, and I thought that was very interesting. That was before I ordered mine.''
    \end{quote}
    
\subsubsection{Learning to Use AI-enabled products}
Participants (P2, P3, P4, P5, P6, P10, and P12) learned more information about their products by reading instruction manuals. They tended to prefer printed manuals over online manuals because it was easier to look back information on printed manuals. However, not every product has a printed manual any more. The situation further aggravates when the product was new and there were no other online sources like YouTube videos to refer to.
\begin{quote}
     P2: ``... I got an iPhone for example uh I thought I would be smart and got the latest possible model so about three years ago I got an iPhone 10 XS and I realized pretty quickly I made a mistake because that was so new there was not much information on the internet on how to do things and if I asked a specific question to get advice they would have it like for an iPhone 7 or an iPhone 8 but they wouldn't have it for an iPhone 10 XS.''
\end{quote}

Many times manuals are not indexed and condensed properly, which makes extracting a piece of information challenging and time-consuming. Moreover a few times, the manuals do not have the information that the interviewees are looking for. The participants also struggle in understanding the language of the manual. For instance, 
\begin{quote}
    P10: ``... I find manuals often don't have enough information...
    and sometimes it's clear that the manual wasn't written by someone whose native language is English, so it took some time to interpret it correctly.''
\end{quote}

\subsubsection{Efforts to Refine Results}
Several participants (P1, P2, P6, P7, and P9) reported that they use voice-assistants for asking factual questions like weather or news. The AI-enabled voice-based systems break when asked for more subjective questions. In situations, when participants looked for a specific output, they tried to interact with the voice-based system by appending a keyword to the question to refine the results. For instance, P7 shares that she wanted to listen a bedtime story ans asked the voice-assistant but the story played had loud background sound, which was not quiet. As a consequence, P7 tried using several keywords back and forth like `quiet' until she got the desired outcome.
\begin{quote}
    P7: ``... I find that I have to add to the keywords. For example, so I may say to her (Alexa) `tell me a bedtime story' or `tell me a quiet bedtime story' so I have to add the word `quiet' to it (Alexa) um she (Alexa) gave me choices and I chose and when I chose I think it was I thought it was quiet but it wasn't quiet and um you know that kind of thing where I need to add additional keywords so that there are maybe five or six key words before she and I reach an agreement on what I want. I might ask for slow jazz, you know, I might ask for jazz music and I would have to ask her for `slow' jazz uh `quiet' jazz `gentle' jazz, whatever, I just keep adding keywords to it until she comes up with what I need.''
\end{quote}
From the above example, which was one of the many instances showed that older adults invested their efforts and tweaked their interaction behavior to refine the voice-assistant's results.

\subsubsection{Recommendation Bubble --- Feedback Loop}
Participants (P6, P8, P9, P12, P13) found personalized recommendations based on a user's online behavior could enhance their entertainment and shopping experience. However, such recommendations create a negative user experiences by putting them in a bubble and hindering them from exploring other diverse content/product on the platform.
For example, participants found video recommendations on YouTube and Netflix extremely restraining as it inundates their platforms with similar content or genre. 
\begin{quote}
    P12: ``... When I start looking at um videos it seems to be um sometimes too much and then sometimes too specific. I'll watch it, I'll tell you for instance, I'll watch an accident connection sequence in YouTube and then I see all these options over and they're just violent violent violent, and that wasn't the reason that I watched this (accident connection sequence video) because it (accident connection sequence video) had a purpose, it was there, but all of a sudden it (recommendations on YouTube) just starts bringing all the stuff here, and it's almost like a shotgun blasted.''
\end{quote}
To overcome this, P6 mentioned that he searched a video in an incognito mode on the browser to avoid the recommendation system from swamping the platform with the specific genre of video.

Similarly, for shopping platforms like Amazon, participants experienced that the recommendation system suggests the user about similar product but those are inaccurate. Other times, these recommendation systems did false advertisements of the product. For instance, P8 received book recommendations based only on the author of the purchased book rather than the topic of the book, which interested her more.
\begin{quote}
    P8: ``... If I was reading a book uh you know if I would you know if I read a book uh about the holocaust, or I bought a book I would get a flood of recommendations for books by the author. I didn't buy the book because of the author, I bought the book because of the topic.''
\end{quote}

\subsection{Challenges in Using AI-enabled Products (RQ2)}
In this section, we discuss the challenges that older adults encountered when using AI-enabled products and the reasons for these hurdles. Participants encountered various hurdles due to the lack of awareness of AI and its perceived usefulness to older adults. Moreover, when AI-enabled products functioned unexpectedly, it made them anxious about their privacy, and ultimately lowered their trust in these products.

\subsubsection{Deviating From Their Expected Behavior}
Our participants (P2, P3, P4, P7, P8, P9, P12, and P14) reported that sometimes their expectations did not match the outcome of an AI-enabled product. People found it frustrating when the AI-enabled product tried to change certain things in a way that the user was not intending. For example, P14 found the voice-assistant in the smartphone annoying when it replaces unfamiliar words with something totally different that the user never intended to use. Moreover, P2 shared that the voice-assistant fails to respond when he is driving and looking for instant responses. The erratic nature of such products puzzles the participants as it does not provide any explanation for their decisions. For example, P12, shares an anecdotal story where he expected the product to justify their decision.
\begin{quote}
    P12: ``When I went to my dentist, Google Map took me down one way and back another way. I don't understand that. Now that I become familiar with the area, I just go back the same way as I came in. When it (Google Map) did that, I got disoriented...I would appreciate the re-routing if it tells me a reason for it, like there's a construction, or an accident.''
\end{quote}

\subsubsection{Over-reliance on AI is Adversely Impacting Decision-making Skills}
The participants (P7, P8, P10, and P12) worried that the growing use of AI in products to enforce viewing of repeated information could replace their intuitive decision-making skills. For example, a few participants (P3, P4, P5, P8, P10, P12) also said that the use of AI to make critical decisions in medicine to do an intrusive procedure or not is agitating. Moreover, they added that the zeal to buy an AI product and trust it is so infectious that people have stopped paying attention to details, like learning who they are building the products for, and do not think through the consequences before depending on such products.
\begin{quote}
    P10: ``... I find the young people to be very reliant, and what I would consider to be over-reliant on technology. So yeah, I think the failure is to not look at this AI as a tool rather than as an end product... AI has learned to distinguish what looks abnormal on a mammogram better than a human eye can do, okay, but abnormal doesn't necessarily mean scary. So the identification of greater abnormality has led to more invasive procedures to do diagnostic procedures.''
\end{quote}

\subsubsection{Pretending to Be Helpful}
Participants (P1, P2, P5, P8, P10, and P12) further talked about AI-enabled customer services in forms of automated telephone trees and chat bots sometimes pretend to be helpful and were intentionally used by companies to trick and prevent them from getting any help. 
\begin{quote}
    P2: ``... I think they intentionally keep it (assistance) difficult to find because they don't want to spend a lot of time with people to do that right.''
\end{quote}

P8 shared became difficult for her to distinguish between the human and the bot interacting with her. To provide a clearer understanding about the challenge we report the experience of P8 below with the AI customer service.
\begin{quote}
    P8: ``... I wanted to know more about my gas bill, so I got on the website there's a chat thing. I thought that the way it (chat) worked was that there was somebody sitting at a laptop and then directly responded to what I typed. But that was not the case. So it was almost as bad as the telephone tree and so I try never to use them.''
\end{quote}

These experiences discourage older adults from using and trusting such AI-enable customer services.

\subsubsection{Data Privacy Threat}
One of the biggest challenges that the older adults (P3, P4, P5, P6, P7, P8, P10, P12, P14, and P15) faced is the threat of keeping their data private and secure because they felt they are more vulnerable and can be taken advantage of. Constant personalized recommendations for products and TV shows used information that they browsed or were asked for, which was an invasion of their privacy (P3, P6, and P12). They (P3, P8, and P10) shared that being unaware of how and who manages the accumulated data make them anxious about using an AI-enabled product. 

\begin{quote}
    P3: ``I think the worry has to do with data accumulation, who is accumulating the data about me as an individual, or about us as a society, and the usage to which that data could be put. I think that that is my only true concern about it [AI] who has control of data.''
\end{quote}

P15, mentions how she turns off the location data every time she is done using a product to avoid any possibility of sharing data unknowingly with the AI-enabled products.
\begin{quote}
    P15: ``... Well the big one is privacy I am a little super crazy about keeping my privacy. So as I said, I turn location off, I close apps as soon as I use them, I turn my computer off when I'm not using it to unplug it, you know, I have all kinds of antiviral and try everything on it, I have ad blockers I don't get ads. It drives my husband nuts he gets ads all the time, but um I really I don't want to be bothered. I don't want that kind of waste of my time, let alone I'm always afraid, well is this some um you know Serbian tracker uh trying to get in. So as I said, I keep lots of privacy blockers on my internet and on my computers, and so i really don't want suggestions from AI unless I ask.''
\end{quote}

\subsubsection{Gender based Expectations from AI}
We observed that different gender-groups respond to AI-enabled products differently.
Female participants (P13, P14, and P15) who live with their partners seem reluctant in setting up the AI-enabled products. 

\begin{quote}
    P13: ``... I don't know why I let my husband handle all the technical stuff, I don't wanna, I can't do it. He said we get a new computer, he sets it up, and I don't do anything.'' 
\end{quote}

P8 and P10 (mentioned about his wife) also mentioned that they and most of her female friends care less about the inner working of a product, and just wanted to know if it could be used easily.
\begin{quote}
    P8 \textit({A female interviewee}): ``... I think for my colleagues and I don't know if it's gender specific, but for me and the women that I have lunch with or socialize with it's a procedure-based thing. We really don't care how it works. I don't care about the underpinnings...what we care is how to access AI what buttons do I push...''
\end{quote}

\subsubsection{Hard-To-Keep-Up with the Changing AI Trend}
Many interviewees (P1, P4, P5, P8, and P9) feel that AI is a growing trend, and it is hard to keep pace with the advancing and ever-changing and complex technology AI.
\begin{quote}
    P1: ``... The other thing is just how rapidly everything gets updated and you have to keep learning it, you know, and you get older, you don't want to learn stuff all the time. It's not that interesting, you want to be able to just use stuff, even like the car. I've got a new car I'd like to be able to drive it without all of this (smart features like conversational agents) stuff and you know it does not have to say, see you every time I get out of the car... It just is all annoying you know life is um more complicated than it used to be, it feels more complicated and more unsafe.'' 
\end{quote}

\subsubsection{Lack of Awareness of AI, and Reasons}
We also found that lack of awareness and knowledge of AI-enabled products put older adults (P1, P2, P4, P5, P7-P9) at a disadvantage to use the product efficiently.
\begin{quote}
    P1: ``... In the communication world there are so many things we can do now and in many cases uh I know I'm not using my um iPhone nearly as effectively as my observation is that my children and certainly our grandchildren uh are taking much more advantage of it. So that type of education is I think important and helpful.''
\end{quote}

The participants (P4, P7, P8, P9, and P10) pointed out two main reasons for their lacking awareness of AI. The first is if an individual does not have a  community such as a workplace or learning centers they miss the opportunity to learn and grow from people about advancing technologies like AI. For example,
\begin{quote}
    P8: ``... I think it's (reason for lacking awareness of AI) not so much age as it is uh what you're doing, if you're not with people using it, doing it, you're out of the loop. There should be some sense of community, like either through workplace, or maybe from uh like someplace, where you see people around you using certain technology and then you leverage that, and then you learn...''
\end{quote}

The second cause mentioned for low awareness of AI in older adults and their apprehensions is because of the negative portrayal of AI-enabled products by media. For instance, P10 says,
\begin{quote}
    P10: ``... There's a lot of suspicion of AI technology...Every time uh Elon Musk's cars kill somebody it makes headlines, but what doesn't make headlines is the number of accidents that are prevented by this technology, which I think is much greater number...'' 
\end{quote}
P1 and P2 also felt the product advertisements fail to convey the usefulness of the products.


\subsection{Older Adults' Perception of AI Technology Based on their Personal Experiences with AI-enabled Products (RQ3)}
\label{subsec:smart}
In this section, we present various themes on older adults' perceptions of AI based on their experiences with the AI-enabled products. A common theme among older adults' perceptions was that AI-enabled products exhibit {\em smartness/intelligence}. However, the perceived smartness was not always positive for example, they consider AI-enabled products as smart threats, which are more dangerous.  

\subsubsection{As an Aid to Save Efforts}
Almost all the participants perceived AI as an aid that helped them accomplish various day-to-day tasks with ease. Continuous advancements in technology such as improvements in voice recognition systems was beneficial in making a phone call or ordering a TV program to watch. Moreover, features like spelling suggestions on smartphones and keyword suggestions on Amazon shopping website minimized the participants' cognitive load while interacting with the product. Navigation applications like Google Maps, Waze made the world more accessible to them. Furthermore, AI-enabled products like Google Search, Alexa enabled them to access desired information instantly. Participants also found AI applications such as smart washer systems, switching on/off security cameras, and driving safety features in cars useful and convenient. 

\begin{quote}
    P14: ``... I like being able to see a print out of the message on my answering machine um I like the fact that I can make a phone call just using my voice, I can order up a television program.''
\end{quote}

\subsubsection{As a Spy to Direct Targeted Advertisements}
Most of the participants (P2, P3, P5, P6, P7, P8, P9, P10, P12, and P14) considered AI as a tool that tracks their online browsing behavior and collects data about them to direct targeted advertisements and steer their data search in a specific direction. They further shared that AI gathers information about their preferences --- their likes and dislikes --- through their computer and phone usage. The advertisements are catered towards the things they have either searched on a browser or have communicated to voice assistants. 

A few participants (P7, P14) also shared that their smartphone and voice-assistant eavesdrop on their conversation happening near them to advertise certain things from the conversation. For instance,
\begin{quote}
    P7: ``... If I have a conversation about something, and I'm not sure if this is what happens but I'll get ads on my Facebook page let's say for things that I've talked about uh with my husband or with friends or on the phone. So I know that she's (Alexa) listening, so that's uh that's the first thing that makes me aware of AI.''
\end{quote}
Participants (P2, P3, P5, P6, P7, P8, P9, P10, P12, and P14) found these experiences annoying and threatening. For example, P6, fears that personalized advertisement recommendations could enable someone to track and harm them. 
\begin{quote}
    P6: ``... I do get concerned when I'll look up something, I'll google something uh a pub a place to go visit, or uh a piece of you know a pair of pants, or something, and then I'll see all kinds of pop-ups about pants or Scotland, you know, I get concerned about that... what if I look for something to buy and then somebody decides oh we're gonna go get all the people who buy that thing and we're gonna do something to them.''
\end{quote}

\subsubsection{ As a Simulator for Human Intelligence}
While a few participants (P2, P7, P8, P12) drew parallels between the AI's intelligence and functioning to the human intelligence and brain to make decisions. They also pointed out AI's desperation to think and act as humans but no matter how many enhancements are made there are always going to be limitations. P7 compares Alexa to 'Pinocchio', a character from the story Pinocchio, who dreams of being a real boy. 
\begin{quote}
    P7: ``... I always think of the story of Pinocchio, well, he always wanted to be a real boy and I think that she wants to be a real girl but she's not a real girl and you just have to accept the fact that AI is going to always have limitations, they (AI) may get better but they may not be able to perfect it.''
\end{quote}
P8 reported that she mistook a Chatbot for an actual human answering their questions until it asked for their feedback objectively (Yes/No) for every answer.

\subsubsection{As a Weapon to Fulfill Malicious Agenda}
Most of the participants (P1-P8, P10, and P11) also perceived AI as tool to fulfil selfish maligning agendas of government, cyber attackers, and big tech companies. The interviewees worried that the government can have an upper hand over their data and can interfere in people's lives. People's personal judgements can be distorted by using AI for disinformation. Furthermore, almost all the interviewees feel AI can be used for wrong reasons like identity theft, scams, and system hacking.
\begin{quote}
    P2: ``... I think that certain uses of AI can be misdirected now. Things like, all this stuff with twitter and whatnot, I don't know to what degree, say for example, the Russians are using AI to figure out you know what people might be sensitive to, to make them doubt that we have good voting systems and you know planting doubt in your head by listening to what people are saying, and then sort of capitalizing on that so all this stuff in the wrong hands be used for doing underhanded things.''
\end{quote}

\subsection{Summary of Interview Findings}
\label{subsubsec:interviewsummary}
Our findings revealed that older adults adopted AI-enabled products through careful considerations like reading consumer reports, recommendations from friends and family. Moreover, in specific situations, they would start using the product if it was gifted or witnessed in use. Since AI is a trending technology, they found it hard to keep pace. Often limited knowledge about AI products is due to lack of exposure to a tech-savvy community and negative portrayal of products in news/media, which hampered them from exploring or using the products. Older adults also encountered several challenges while using AI-enabled products like a threat to their privacy, over-reliance on AI impacting decision-making skills, and AI outcomes outwitting them. We further examined older adults' perceptions of AI based on their experiences. We found the quality of older adults' experiences with AI-enabled products formed their perception of AI. In many cases, older adults felt AI-enabled products assisted them in their day-to-day tasks. In contrast, at the same time, bad experiences like targeted advertisements, scams, and frustrations interacting with conversational agents concerned them.~Therefore, in the next section, we discuss different design considerations for AI-enabled products that could enrich older adults' product experiences.

\section{Discussion}
\label{sec:discussion}

\subsection{Older Adults' Experiences in Using AI and Perceptions of AI}


We found that older adults' perceptions of AI were based on {\em perceived smartness} of AI (see Section~\ref{subsec:smart}), which were two-fold: i) \textit{AI as an assistance}, and ii) \textit{AI as a nuisance or a threat}. They shared various instances explaining how AI helped them in their day-to-day activities; at the same time, they also reported their concerns about AI as a threat to their privacy and a tool to accomplish vicious agendas. A previous study by Shin and Lee~\cite{shin2014effects} discussed the product's perceived smartness played a critical role in influencing a customer's understanding of the product's advantages and user acceptance. Our study found that an AI-enabled product's perceived smartness not only helped older adults identify its benefits but also its disadvantages.
Moreover, older adults' motivations to accept a product depends on its perceived usefulness, ease of use (usability), recommendation from close ones, and assistance in using a product, which echo prior studies~\cite{lee2015perspective,wang2011older,xie2007information}. We further discovered that older adults' introduction to latest AI-enabled products is dependent on knowledge gained about the usefulness of the product from friends/family, availability of the product in the market, consumer reports, and public opinions about the products expressed in reviews on online shopping websites.

Our study found that females were disinclined towards learning about the inner-functionalities of the AI-enabled products but were excited about using them, and females with partners often depended on their partners to set-up AI-enabled products. Since we do not have insights from other genders, our study does not provide a generalized understanding of whether and how gender might play a role in the adoption of AI-enabled products among older adults and calls for further research investigation. 
Supplementary to this result, the difference in technology adoption and confidence between gender groups was documented in past studies~\cite{goswami2015gender,yau2012gender}. 

We found that factors like threats of data privacy, trusting AI-enabled products when the outcomes are unexplainable make older adults anxious in using AI-enabled products, which was also shown to be true for younger users~\cite{ostrom2019customer}. Our participants felt some AI products, such as AI customer-service, were designed to make people feel being served by a real person but were not instrumental to assist. Such feelings about AI negatively affected older adults' trust in AI-enabled products.

\subsection{Design Considerations}
In the findings (Section~\ref{sec:interviewfindings}), we presented varied experiences of older adults with the AI-enabled products. Based on older adults' encounters and challenges in using AI-enabled products, we discuss the design considerations to improve perceived usefulness and trust in AI-enabled products for older adults in light of the past work.


\subsubsection{\bfseries Implicit Feedback Based on Users' Interaction to Improve Efficiency of the AI-enabled Products}
We discovered that older adults feel puzzled when the outcomes of AI-enabled products do not match their expectations. For an older adult to fully embrace AI technology, the outcome of the products must match users’ expectations, or it should effectively answer what they were looking for. Many participants (P2, P4, P7, P9, P10, P11, and P12) felt that voice-assistants and automated telephone trees eluded from providing information that they wanted to learn. Moreover, those AI-enabled products (voice-assistants and automated telephone trees) absorbed older adults’ energy in futile back and forth interactions with no relevant outcomes, ruining their experiences. For instance, participants (P2, P7, and P9) shared that the voice-assistants defy users’ commands and outwit them, such as referring to internet pages when asked about defining a term and narrating a noisy bedtime story when the user wanted to listen to a calming bedtime story.

Therefore, to improve the efficiency of AI-enabled products, the products should carefully monitor users’ interactions to assess if the outcome was desirable. The products should collect feedback implicitly on their outcomes from the users based on the user interaction cues such as, `repeated queries are asked using different keywords,’ `time spent interacting with the product without the desired outcome,’ or `turning off the product.’ For instance, if a user struggles to get the desired information using an automated telephone tree then instantly let the product ask if they need help and provide them with a suitable response or human assistance.

\subsubsection{\bfseries Explicit Consent to Collect Data When a User Interacts with AI-enabled Products}
One of the main concerns shown by all the participants in using AI-enabled products is the threat to their data privacy. The problem further aggravates when a user is not even aware of their data stealthily being used in an AI-enabled product. 

To overcome the problem of data privacy, the products should \textit{explicitly inform the users about the current data being tracked and explain why and how it will be used.} Moreover, provide the user control to opt-in sharing their data.

\subsubsection{\bfseries Periodic Walk-throughs and Data Usage Summary to Manage Privacy Settings}
Another big issue that our participants (P3, P7, P13, P14, and P15) struggle with is managing their privacy settings in an AI-enabled product. This becomes an even bigger problem because they are less tech-savvy as compared to younger user groups. Therefore, to address this issue, AI-enabled products should provide: i) walk-throughs and reminders of privacy settings to inform and educate users about it and ii) user's data usage summary report periodically that makes them aware and keep them in the know. 
Recently, Nora and Mentis~\cite{Nora2021BuildingforWe} showed that pairing older adults up to deal with privacy and security information together could be another viable approach.

\subsubsection{\bfseries Enable Users to Activate or Deactivate Listening Mode in AI-enabled Products}
Many participants (P4, P9, P13, P14, and P15) hesitated to casually converse around AI-enabled products, specifically voice-assistants, because they felt the device constantly eavesdropped on their conversations. They suspected this phenomenon mainly because of two reasons: i) targeted advertisements about products/services that were merely brought up in their conversations, which happened near the device, and ii) whenever the voice-assistants such as Alexa, were talked about in a conversation, voice-assistants would respond (we also observed this phenomenon while interviewing the participant.)

Thus, users should be enabled with the option to activate/deactivate the listening mode in an AI-enabled product. The product in a deactivated listening mode would not respond when they are talked about. To further alleviate the privacy apprehensions of \textit{users, provide them control to activate or deactivate the listening mode whenever they want.} Moreover, deactivate product's listening mode automatically in certain circumstances such as i) time of the day, ii) work and non-work hours, iii) location (home and office), iv) public gatherings, and v) if any electronic device (television saying 'Alexa' or 'Google') is already in use.

\subsubsection{\bfseries Choice to Learn/Skip the Inner-workings of AI-enabled Products}
It is worth noting that controlling the settings of AI-enabled products often demands some understanding or mental model of the underlying AI. In a few instances, our participants (P8, P10, P13, P14, and P15) indicated their reluctance to learn the underpinnings of a product and wanted the product to operate without them worrying about the inner-functionalities. Therefore, they wanted an \textit{alternative to learning or skipping the underpinnings of an AI-enabled product}, which involved understanding if a product uses their data and how. Although this problem does not directly stem from the products that participants used, however, participants wanted this feature in an {\em explainable AI-enabled product}.

\subsubsection{\bfseries Break the Feedback Loop: Make the Explore Option Always Visible}
Our participants (P6 and P12) indicated that recommendations on video platforms inundated an older adult's video platform with repetitive content/genre, which they found annoying, as it limited their opportunity to engage with new content. Prior studies~\cite{fields2011analysis,wilhelm2018practical} have also discussed the problem of stale and unwanted recommendations, but this issue persists in the products.

To break the feedback loop, there should be a \textit{visible option, which allows users to explore more than what is recommended/presented.} For instance, YouTube has a specific tab 'Explore,' letting users discover content beyond recommendations.


\subsubsection{\bfseries Interactive Shopping Recommendations}
Participants P3 and P8 expressed their dissatisfaction with the shopping recommendations suggested by the shopping portal. P8 shared that the proposed book recommendations were based on the author of the purchased book and not on the topic. The user did not have a way to notify that the shopping recommendations were not helpful when looking for a book on the same topic rather than the same author.
To address the issue: i) there should be an option for the user to inform that a specific recommendation is not relevant, ii) recommendations should be based on multiple attributes of the products instead of one, for example, the above instance only considered the author.

\subsubsection{\bfseries Well-structured Printed Manuals for AI-enabled Product}
Participants (P2, P4, P5, P8, and P10), specifically aged 70 or above, preferred printed manuals with well-structured senior-friendly descriptions over digital resources, which is in line with previous studies~\cite{champley2008preliminary,meyer1997age,tsai2012older}. These days, they mentioned that many trending products like iPhones and voice-assistants do not provide comprehensive user manuals outlining detailed processes to interact with the product. For instance, P2 said that he bought a trending smartphone, thinking it was the latest product in the market, but since it came with no user manual, it became challenging for him to interact or use the phone to its full potential, and as a result, he refrained from using the smartphone. Besides, navigating an online user manual is another difficulty for them, and sometimes they do not even get the desired information. P4 shared a similar experience; when she bought Alexa, it was confusing for her to differentiate between the blinking colors and the device randomly lighting up without having an access to a user manual describing the interaction details. Therefore, our participants desired affordable cues in AI-enabled products to cater to aging needs. To cater to older adults' aging needs and minimize cognitive load, AI-enabled products should incorporate gradual disclosure and explanation of features~\cite{druga2017hey,long2020ai}. 

Moreover, the AI-enabled products should \textit{ensure to provide well-written senior-friendly documentation} explaining: i) how to set-up the device, ii) categorization of user manual depending on the various tech-savviness levels of the user --- how to interact with the device at a bare minimum, intermediate, and advance level, iii) ways to address plausible device errors, and iv)  how to seek assistance when required. Fan and Truong proposed design guidelines for creating senior-friendly instructions for non-AI products ~\cite{fan2018guidelines}, which can be extended for AI-enabled products.

\section{Limitations and Future Work}
 \label{sec:limitations}


To our knowledge, our work is a first step towards understanding older adults' personal experiences and perceptions of AI-enabled everyday products. Our design recommendations provide a starting point for designers and researchers to think about ways to make AI-enable everyday products more accessible to older adults. Nonetheless, there are some important limitations. In our study, we did not provide our participants with a formal definition of AI but instead provided a list of reasonably representative sample AI-enabled products to elicit older adults' authentic understandings of AI and AI-enabled products. One potential limitation is that our findings are based on older adults' subjective experiences and perceptions rather than an objective measure of their AI knowledge. Moreover, as AI technology continues to be embedded into more products, the sample AI-enabled products used in our work might not represent all possible AI-enabled products. Future work should extend our work to investigate older adults' objective AI knowledge and their experiences with a wider range of AI-enabled products.  

Our participants were from the North-American region, thereby providing us with the outlook and perception of older adults living in developed areas. It will be interesting to investigate the impact of AI advancements in developing regions to learn whether and to what extent older adults' experiences and perceptions of AI differ from our findings. Additionally, it will be interesting to compare older adults' and younger adults' perceptions of AI.

\section{Conclusion}
\label{sec:conclusion}

We have presented a survey study and a qualitative interview study to understand older adults' perceptions and knowledge of AI and their experiences in using AI-enabled products. We found that older adults identified the use of AI in products by associating \emph{smartness} in the AI-enabled products. However, they also perceived AI as an aid, a tool that simulates human intelligence but it would never become as efficient as a human, and a spy to direct targeted advertisements. 
We identified the challenges that they encountered when using AI-enabled products. They felt that AI used in the product misaligned with their expected behavior, perceived it as a  trend that is hard-to-keep-up with, a threat to privacy, and a technology we over-rely on without having any controls on it.
Based on the older adults' experiences with the AI-enabled products, we presented design considerations for designers and researchers to make AI-enabled products more usable and accessible for older adults.


\bibliographystyle{ACM-Reference-Format}
\bibliography{main}

\end{document}